\RequirePackage{fixltx2e}
\documentclass[oneside,english]{autart}
\usepackage[T1]{fontenc}
\usepackage[latin9]{inputenc}
\usepackage[a4paper]{geometry}
\usepackage{verbatim}
\usepackage{amsmath}
\usepackage{amssymb}
\usepackage{stackrel}
\usepackage{graphicx}

\makeatletter

\providecommand{\tabularnewline}{\\}

	\makeatletter
	\def\old@comma{,}
	\catcode`\,=13
	\def,{%
  		\ifmmode%
    		\old@comma\discretionary{}{}{}%
  		\else%
    		\old@comma%
  		\fi%
	}
	\makeatother
\usepackage{cite}
\usepackage{multirow} 
\usepackage{accents}
\usepackage[english]{babel}
\usepackage{amsfonts}
\usepackage{amssymb}
\allowdisplaybreaks
\makeatother

\usepackage{babel}
\begin{document}
\begin{frontmatter}
\title{Lyapunov-Based Physics-Informed Deep Neural Networks with Skew Symmetry
Considerations\thanksref{grants}}
\thanks[grants]{This research is supported in part by the National Science Foundation
Graduate Research Fellowship under Grant No. DGE-2236414, AFRL grant
FA8651-24-1-0018, and AFOSR grant FA9550-19-1-0169. Any opinions,
findings, and conclusions or recommendations expressed in this material
are those of the author(s) and do not necessarily reflect the views
of the sponsoring agency.}
\thanks{}\author[UF]{Rebecca G. Hart,}
\author[AFRL]{Wanjiku A. Makumi,}
\author[UF]{Rushikesh Kamalapurkar,}
\author[UF]{Warren E. Dixon}

\address[UF]{Department of Mechanical and Aerospace Engineering, University of Florida, Gainesville, FL 32611 USA. Email: {rebecca.hart, rkamalapurkar, wdixon}@ufl.edu.} 
\address[AFRL]{Munitions Directorate, Air Force Research Laboratory, Eglin AFB, FL 32542 USA. Email: wanjiku.makumi@us.af.mil.}
\begin{abstract}
Deep neural networks (DNNs) are powerful black-box function approximators
which have been shown to yield improved performance compared to traditional
neural network (NN) architectures. However, black-box algorithms do
not incorporate known physics of the system and can yield results
which are physically implausible. Physics-informed neural networks
(PINNs) have grown in popularity due to their ability to leverage
known physical principles in the learning process which has been empirically
shown to improve performance compared to traditional black-box methods.
This paper introduces the first physics-informed DNN controller for
an Euler-Lagrange dynamic system where the adaptation laws are designed
using a Lyapunov-based stability analysis to account for the skew-symmetry
property of the inertia matrix and centripetal-Coriolis matrix. A
Lyapunov-based stability analysis is provided to guarantee asymptotic
convergence of the tracking error and the skew-symmetric prediction
error. Simulations indicate that the developed update law demonstrates
improvement in individual and overall function approximation capabilities
when compared to a physics-informed adaptation law which does not
incorporate knowledge of system symmetries.
\end{abstract}
\end{frontmatter}

\section{Introduction}

Learning methods such as neural networks (NNs) have become popular
black-box estimators of system dynamics \cite{Patil.Le.ea2022,Lewis1996b,Basyal.Ting2024,Bansal.Akametalu.ea2016,Makumi.Bell.ea23,Philor.Makumi.ea2024,Greene.Bell.ea.2023}
due to their ability to compensate for unknown dynamics. Compared
to shallow NNs, deep neural networks (DNNs) have been shown to be
more expressive \cite{Rolnick.Tegmark2018}. Black-box estimators
like NNs and DNNs use offline optimization techniques which focus
on fitting the output to observed data during training. One inherent
limitation is the dependence on sufficiently rich datasets for training
and validation, along with the lack of guarantees for adherence to
known physical laws. Ensuring estimates satisfy known properties
of the system reduces the search space of desirable approximations
and can provide additional interpretability insights. Physics-informed
learning which embeds established physical principles into the learning
architecture has been explored in results such as \cite{Raissi.Perdikaris.ea2019,Karniadakis2021,Cranmer.Greydanus.ea2020,Lutter2019,Hart.Griffis.ea2024,Hart.patil.ea2023}.
As indicated in \cite{Karniadakis2021}, incorporating constraints
and governing equations enables physics-informed neural networks (PINNs)
to generate predictions that have improved accuracy, physical plausibility,
and performance for exploratory or generalization tasks where black-box
architectures often struggle.

A popular way to incorporate known properties is to include symmetry
constraints, which play a role in many physical systems. Various symmetry
types including rotational, translational, even-odd, and time-reversal
symmetries have been explored for NN learning in results such as \cite{Wang.Walters.2021,Anderson.Hy2019,Mattheakis.Protopapas.2020,Huh.Yang.2020}.
In systems governed by Lagrangian and gyroscopic dynamics, such as
robotics and spacecraft mechanics, skew-symmetric properties naturally
emerge \cite{Lian.Wang.1998}. Knowledge of these properties is used
in \cite{Tian.Livescu2021}, which uses a NN and offline optimization
techniques to learn a scaling factor for the symmetric and skew-symmetric
integrity bases in flow models. The work in \cite{Rozema.Verstappen2014}
leverages skew-symmetry properties in a discretization scheme, demonstrating
that the physical conservation properties of convective terms can
be maintained by preserving their skew-symmetric structure. While
both approaches focus on modeling nonlinear systems for simulation
or analysis, they do not directly address control applications. The
result in \cite{Lutter2019} embeds the Euler-Lagrange equation into
a NN framework and uses offline optimization to minimize violations
of the Lagrangian, which implicitly preserves the skew-symmetric structure.
The resulting estimated model is then used as a feedforward term in
a PD control scheme, demonstrating an application of physics-informed
learning in control.

To overcome the limitations of offline training, Lyapunov-based (Lb-)
DNN adaptation laws were developed in \cite{Patil.Le.ea2022} to enable
real-time weight updates of DNNs while providing stability guarantees.
This real-time learning capability is particularly beneficial for
nonlinear systems experiencing changes to system dynamics, uncertain
effects, or unmodeled phenomena such as viscoelastic or electromagnetic
forces. However, approaches which use real-time adaptation such as
\cite{Makumi.Bell.ea23,Joshi.Virdi.ea2020} and \cite{Makumi.Bell.ea2024}
are black-box approaches which disregard known system properties.
Other Lb-DNN developments such as \cite{Hart.Griffis.ea2024} and
\cite{Hart.patil.ea2023} have developed a physics-informed framework
using the known structure of the Euler-Lagrange equation and use individual
function approximators to estimate the unknown matrices. However,
the developed PINN architectures use tracking-error-based adaptation
laws and do not include any additional physics-based constraints on
the parameter adaptation.

Motivated by the need to incorporate both dynamic structure and system
properties, this is the first work which introduces a physics-informed
regularization into the real-time stability driven adaptation law.
The developed approach, termed Skew-Symmetric Lyapunov-based PINN
(SS-LbPINN), achieves real-time adaptation of unknown model terms
and regularizes the weight adaptation of the inertial and centripetal-Coriolis
matrices towards their known skew-symmetric property. The key technical
innovation required to embed the symmetry properties in the adaptation
is the development of a skew-symmetry based prediction error. The
analytical form, in particular, clarifies the theoretical impact of
the physics-informed prediction error by directly relating it to individual
parameter estimation errors within the skew-symmetric matrices.  A
Lyapunov-based stability analysis is performed to design analytical
real-time update laws that are included within a controller to ensure
asymptotic error convergence for both a desired trajectory tracking
error and the skew-symmetric prediction error. Simulations were performed
on an Euler-Lagrange system and yielded a root mean square (RMS) tracking
error of 0.0195 rad. The incorporation of the skew-symmetric prediction
error resulted in a 19.87\% improvement in overall function approximation
capabilities and 4.84\%, 21.78\%, and 2.52\% improvement in the approximation
of the unknown inertia, centripetal-Coriolis, and friction matrices,
respectively, compared to the baseline method in \cite{Hart.patil.ea2023}.

\section{Problem Formulation}

\subsection{Notation and Preliminaries}

The space of essentially bounded Lebesgue measurable functions is
denoted by $\mathcal{L}_{\infty}$. Given $A\triangleq\left[a_{j,i}\right]\in\mathbb{R}^{n\times m},$
$\text{\text{vec}}(A)\triangleq\left[a_{1,1},\ldots,a_{n,1},\ldots,a_{1,m},\ldots,a_{n,m}\right]^{\top}$.
The Kronecker product is denoted by $\otimes$. Given any $A\in\mathbb{R}^{n\times m}$,
$B\in\mathbb{R}^{m\times p}$, and $C\in\mathbb{R}^{p\times r}$,
$\mathrm{vec}\left(ABC\right)=\left(C^{\top}\otimes A\right)\mathrm{vec}\left(B\right)$.
The notation $\lambda_{\text{min}}(A)$ and $\lambda_{\text{max}}(A)$
denote the minimum and maximum eigenvalues of a matrix, respectively.
The signum function, denoted by $\text{sgn}(D)$ where $D\in\mathbb{R}^{1}$,
is defined as $\text{sgn}(D)\triangleq1$ if $D>0$, $\text{sgn}(D)\triangleq0$
if $D=0$, and $\text{sgn}(D)\triangleq-1$ if $D<0$. For a vector
$\boldsymbol{D}\in\mathbb{R}^{n}$ defined as $\boldsymbol{D}\triangleq\left[D_{1},D_{2},\ldots,D_{n}\right]^{\top}$the
sign function is applied element-wise such that $\text{sgn}(\boldsymbol{D})\triangleq\left[\text{sgn}(D_{1}),\ \text{sgn}(D_{2}),\ldots,\ \text{sgn}(D_{n})\right]^{\top}.$
The notation $\overset{\text{a.e.}}{(\cdot)}$ denotes the relation
$(\cdot)$ holds almost everywhere (a.e.). The notation $\text{K}[\cdot]$
denotes the Filippov's regularization $[\cdot]$ \cite{Paden1987}.
The right-to-left matrix product operator is represented by $\stackrel{\curvearrowleft}{\prod}$,
i.e., $\stackrel{\curvearrowleft}{\stackrel[p=1]{m}{\prod}}A_{p}\triangleq A_{m}\ldots A_{2}A_{1}$
where $\stackrel{\curvearrowleft}{\stackrel[p=a]{m}{\prod}}A_{p}\triangleq1$,
if $a>m$. The identity matrix of size $n\times n$ is denoted by
$I_{n}$. Given functions $f$ and $g$, function composition is
denoted by $\circ$, where $\left(f\circ g\right)(x)\triangleq f(g(x)).$

\subsection{Deep Neural Network (DNN) Model\label{subsec:DNNModel} }

For simplicity, the following development considers a fully-connected
DNN, though the control and adaptation laws can be generalized to
any other architecture\footnote{For illustrations of other architectures the reader is referred to
\cite{Patil.Le.ea.2022,Griffis.Patil.ea23_2,Hart.Griffis.ea2024}.} $\Phi_{i}$. In this paper, physics information is encoded in the
adaptive controller by employing a family of DNNs that model different
parts of the dynamics, such as inertia, centripetal-Coriolis terms,
gravitational effects, and friction. The family of DNNs is denoted
by $\left\{ \Phi_{i}\right\} _{i\in\mathcal{I}}$ where $\mathcal{I}$
is a finite index set. For the $i^{th}$ network, let $x_{i}\in\Omega_{i}$
denote the input to the DNN, and $\Omega_{i}\subset\mathbb{R}^{L_{in,i}}$
denote a known compact set where $L_{in,i}\in\mathbb{R}_{>0}$. The
augmented input which accounts for bias terms is denoted by $x_{a,i}\triangleq\begin{bmatrix}x_{i}^{\top},1\end{bmatrix}^{\top}.$
Let $k_{i}\in\mathbb{Z}_{>0}$ denote the total number of hidden layers,
$v_{j,i}\in\mathbb{R}^{L_{j,i}\times L_{j+1,i}}$ denote the matrix
of weights and biases, where $L_{j,i}\in\mathbb{Z}_{>0}$ denotes
the number of nodes in the $j^{th}$ hidden layer for all $j\in\left\{ 0,\dots,k_{i}\right\} $,
$L_{0,i}\triangleq L_{in,i}+1$, $L_{k+1,i}=L_{out,i}$ and $L_{out,i}\in\mathbb{Z}_{>0}$
is the size of the output, for all $i\in\mathcal{I}$. The vector
of DNN parameters (i.e., weights and bias terms) is denoted $\theta_{i}\triangleq\left[\text{\text{vec}}(v_{0_{i}})^{\top},\ldots,\text{\text{vec}}(v_{k_{i}})^{\top}\right]^{\top}$
$\in\mathbb{R}^{\varkappa_{i}}$, where $\varkappa_{i}\triangleq\sum_{j=0}^{k_{i}}L_{j,i}L_{j+1,i}$.
Additionally, a bounded parameter search space $\Theta_{i}\triangleq\left\{ \theta_{i}\in\mathbb{R}^{\varkappa_{i}}:\left\Vert \theta_{i}\right\Vert \leq\bar{\theta}_{i}\right\} $
is considered where $\bar{\theta}_{i}\in\mathbb{R}_{>0}$ is a user-selected
bound. The individual fully-connected feedforward DNN $\Phi_{i}\left(x_{i},\theta_{i}\right)\in\mathbb{\mathbb{R}}^{L_{out,i}}$
is defined for all $i\in\mathcal{I}$ as $\Phi_{k_{i},i}\triangleq\Phi_{i}(x_{i},\theta_{i})$,
where $\Phi_{j,i}\in\mathbb{R}^{L_{j+1}}$ for $j=0,\dots,k_{i}$
is defined using the recursive relation
\begin{equation}
\Phi_{j,i}\triangleq\begin{cases}
v_{j,i}^{\top}\phi_{j,i}(\Phi_{j-1,i}), & j\in\{1,\ldots,k_{i}\},\\
v_{0,i}^{\top}x_{a,i} & j=0.
\end{cases}\label{eq:Phij_DNN}
\end{equation}
The vector of smooth activation functions at the $j^{\text{th}}$
layer is denoted by $\phi_{j,i}(y)\in\mathbb{R}^{L_{j,i}}$, for all
$j\in\left\{ 1,\ldots,k_{i}\right\} $ and $i\in\mathcal{I}$. The
time derivative of the DNN, represented by \textbf{$\frac{d}{dt}\left(\Phi_{i}(x_{i},\theta_{i})\right)\in\mathbb{R}^{L_{out,i}}$},
follows from the chain rule applied to (\ref{eq:Phij_DNN}), accounting
for the change in both $x_{i}$ and $\theta_{i}$ over time for all
$i\in\mathcal{I}$. The Jacobian of the activation function with respect
to the state at the $j^{th}$ layer, denoted by $\phi_{j,i}^{\prime}:\mathbb{R}^{L_{j,i}}\rightarrow\mathbb{R}^{L_{j,i}\times L_{j,i}}$,
is defined as $\phi_{j,i}^{\prime}(y)\triangleq\frac{\partial}{\partial y}\phi_{j,i}(y)$,
for all $y\in\mathbb{R}^{L_{j,i}}$ and $i\in\mathcal{I}$. Let the
Jacobian of the DNN with respect to the weights be denoted by $\Phi^{\prime}(x_{i},\hat{\theta}_{i})$
which can be represented as $\Phi_{i}^{\prime}\triangleq\left[\Phi_{0,i}^{\prime},\ldots,\Phi_{j,i}^{\prime}\right]\in\mathbb{R}^{L_{out,i}\times\varkappa_{i}}$,\textbf{
}where $\Phi_{j,i}^{\prime}\triangleq\frac{\partial\Phi_{j,i}\left(x_{i},\hat{\theta}_{i}\right)}{\partial\hat{\theta}_{i}}\in\mathbb{R}^{L_{out,i}\times L_{j+1,i}}$,
for all $j\in\left\{ 0,\ldots,k_{i}\right\} $ and $i\in\mathcal{I}$.
Using (\ref{eq:Phij_DNN}), the chain rule, and properties of the
vectorization operator, the Jacobians $\Phi_{0,i}^{\prime}$ and $\Phi_{j,i}^{\prime}$
can be expressed as \cite{Hart.patil.ea2023}
\begin{align}
\Phi_{0,i}^{\prime} & \triangleq\left(\stackrel{\curvearrowleft}{\prod_{l=1}^{k_{i}}}\hat{v}_{l,i}^{\top}\hat{\phi}_{l,i}^{\prime}\right)\left(I_{L_{1,i}}\otimes x_{a,i}^{\top}\right),\nonumber \\
\Phi_{j,i}^{\prime} & \triangleq\left(\stackrel{\curvearrowleft}{\prod_{l=j+1,i}^{k_{i}}}\hat{v}_{l,i}^{\top}\hat{\phi}_{l,i}^{\prime}\right)\left(I_{L_{j+1,i}}\otimes\hat{\phi}_{j,i}^{\top}\right),\label{eq:JacobianDNN}
\end{align}
for all $j\in\left\{ 1,\ldots,k_{i}\right\} \text{ and }i\in\mathcal{I}$.
The time derivative of the Jacobian is represented as $\frac{d}{dt}\Phi_{i}^{\prime}\triangleq\left[\frac{d}{dt}\Phi_{0,i}^{\prime},\dots,\frac{d}{dt}\Phi_{j,i}^{\prime}\right]\in\mathbb{R}^{L_{out,i}\times\varkappa_{i}}$
where $\frac{d}{dt}\Phi_{j,i}^{\prime}\triangleq\frac{d}{dt}\left(\frac{\partial\Phi_{j,i}(x_{i},\hat{\theta}_{i})}{\partial\hat{\theta}_{i}}\right)\in\mathbb{R}^{L_{out,i}\times L_{j+1,i}}$
where $\frac{d}{dt}\Phi_{0,i}^{\prime}$ and $\frac{d}{dt}\Phi_{j,i}^{\prime}$
can be obtained by applying chain rule to (\ref{eq:JacobianDNN})
for all $j\in\left\{ 0,\dots k_{i}\right\} $ and $i\in\mathcal{I}$.
\begin{assum}
\label{assm:activation bounds} For each $j\in\left\{ 0,\ldots,k\right\} $,
the activation function $\phi_{j}$, its Jacobian $\phi_{j}^{\prime}$,
Hessian $\phi_{j}^{\prime\prime}\left(y\right)\triangleq\frac{\partial^{2}}{\partial y^{2}}\phi_{j}\left(y\right)$,
and third-order tensor $\phi^{\prime\prime\prime}(y)\triangleq\frac{\partial^{3}}{\partial y^{3}}\phi_{j}(y)$
 are bounded as
\begin{eqnarray}
\left\Vert \phi_{j}\left(y\right)\right\Vert \leq\mathfrak{a}_{0}, &  & \left\Vert \phi_{j}^{\prime\prime}\left(y\right)\right\Vert \leq\mathfrak{c}_{0},\nonumber \\
\left\Vert \phi_{j}^{\prime}\left(y\right)\right\Vert \leq\mathfrak{b}_{0}, &  & \text{\ensuremath{\left\Vert \phi_{j}^{\prime\prime\prime}\left(y\right)\right\Vert \leq\mathfrak{d}_{0}},}\text{}\label{eq:Activation Bounds}
\end{eqnarray}
where $\mathfrak{a}_{0},\mathfrak{b}_{0},\mathfrak{c}_{0},\mathfrak{d}_{0}\in\mathbb{R}_{\geq0}$
are known constants.
\end{assum}
\begin{rem}
\label{rem:activation bounds} Most activation functions used in practice
satisfy Assumption \ref{assm:activation bounds}. Specifically, sigmoidal
activation functions (e.g., logistic function, hyperbolic tangent,
etc.) have $\left\Vert \phi_{j}\left(y\right)\right\Vert $, $\left\Vert \phi_{j}^{\prime}\left(y\right)\right\Vert $,
$\left\Vert \phi_{j}^{\prime\prime}\left(y\right)\right\Vert $, and
$\left\Vert \phi_{j}^{\prime\prime\prime}\left(y\right)\right\Vert $
bounded uniformly by constants. 
\end{rem}

\subsection{Model Dynamics and Control Objective}

Consider an uncertain Euler-Lagrange system modeled as
\begin{equation}
M(q)\ddot{q}+C(q,\dot{q})\dot{q}+G(q)+F\left(\dot{q}\right)=\tau\left(t\right),\label{eq:dynamics}
\end{equation}
where $q,\dot{q},\ddot{q}\in\mathbb{R}^{n}$ denote the generalized
position, velocity, and acceleration, respectively. The system dynamics
are characterized by the unknown generalized inertia matrix, generalized
centripetal-Coriolis effects, generalized potential forces, generalized
dissipation effects, and the subsequently designed control input which
are denoted by $M\in\mathbb{R}^{n\times n},$ $C\in\mathbb{R}^{n\times n}$,
$G\in\mathbb{R}^{n},$ $F\in\mathbb{R}^{n}$, and $\tau\in\mathbb{R}^{n}$,
respectively. The system in (\ref{eq:dynamics}) satisfies following
properties \cite[Sec. 2.3]{Ortega1998}.\linebreak{}

\begin{prope}
\label{Prop:PDInertia}The inertia matrix $M(q)$, satisfies $m_{1}\left\Vert \zeta\right\Vert ^{2}\leq\zeta^{\top}M(q)\zeta\leq m_{2}\left\Vert \zeta\right\Vert ^{2}$
for all $q\in\mathbb{R}^{n}$, where $m_{1},m_{2}\in\mathbb{R}_{>0}$
denote known constants and $\zeta$ represents a subsequently defined
state vector.
\end{prope}
\begin{prope}
\label{Prop:SkewSymmetric}The time-derivative of the inertia matrix
and centripetal-Coriolis matrix satisfy the skew-symmetry relation,
$\xi^{\top}(\dot{M}(q)-2C(q,\dot{q}))\xi=0$, for all $q,\dot{q},\xi\in\mathbb{R}^{n}$.
\end{prope}
The tracking control objective is to design a controller to asymptotically
track a user-defined, time-varying desired trajectory, $q_{d}\in\mathbb{R}^{n}$,
which is designed to be sufficiently smooth such that $q_{d}(t),\dot{q}_{d}(t),\ddot{q}_{d}(t)\in\mathcal{Q},\ \text{for all }t\in\mathbb{R}_{\geq0}$,
where $\mathcal{Q}\subseteq\mathbb{\mathbb{R}}^{n}$ denotes a known
compact set. To quantify the tracking objective, the tracking error
$e\in\mathbb{R}^{n}$ and auxiliary tracking error $r\in\mathbb{R}^{n}$
are defined as
\begin{align}
e & \triangleq q-q_{d}, & r\triangleq\dot{e}+\alpha e,\label{eq:res_error}
\end{align}
respectively, where $\alpha\in\mathbb{R}_{>0}$ denotes a user-selected
constant control gain. Quantitatively, the tracking objective is to
ensure $\left\Vert e(t)\right\Vert \to0$ and $\left\Vert r(t)\right\Vert \rightarrow0$
as $t\to\infty$.  Using (\ref{eq:dynamics})-(\ref{eq:res_error}),
the open-loop dynamics for $r$ can be determined as
\begin{align}
M(q)\dot{r} & =\tau-M(q)\left(\ddot{q}_{d}-\alpha\dot{e}\right)-C(q,\dot{q})\left(\dot{q}_{d}-\alpha e\right)\nonumber \\
 & \quad-G(q)-F\left(\dot{q}\right)-C(q,\dot{q})r.\label{eq:CLES1}
\end{align}

\subsection{Adaptive Physics-Informed Control Development}

Motivated by the desire to leverage known system information, the
developed controller uses the known structure of the dynamics in (\ref{eq:dynamics})
by modeling the unknown terms using individual DNNs similar to \cite{Hart.Griffis.ea2024}
and \cite{Hart.patil.ea2023}. This formulation provides insight
into the specific inputs to the DNNs, ensuring a more informed learning
process and allows for the development of a skew-symmetric prediction
error due to the availability of individual estimates of the skew-symmetric
components. To ensure the output of the DNN is the appropriate dimension,
vectorized forms of $M(q)$ and $C(q,\dot{q})$ are used along with
properties of the Kronecker product. The use of vectorized versions
of $M(q)$ and $C(q,\dot{q})$ facilitate the subsequent stability
analysis and eliminate the need to unvectorize the DNN outputs. 

Consider the family of DNNs described in Section \ref{subsec:DNNModel}
where $\mathcal{I}\triangleq\left\{ M,C,G,F\right\} $, and let $\mathcal{F}\triangleq\left\{ \text{vec}\left(M(q)\right),\text{vec}\left(C(q,\dot{q})\right),G(q),F(\dot{q})\right\} $
where $\mathcal{F}_{i}$ represents the corresponding vector in $\mathcal{F}$
for the $i^{th}$ element in $\mathcal{I}$. The universal function
approximation theorem \cite[Theorem 3.1]{Kidger.Lyons2020} states
that the function space of DNNs is dense in the space of continuous
functions $\mathcal{C}\left(\Omega_{i}\right)$ for $i\in\mathcal{I}$.
As a result, there exists a DNN $\Phi_{i}(x_{i},\theta_{i}^{*})$
and corresponding parameters $\theta_{i}^{*}\in\mathbb{R}^{\varkappa_{i}}$
such that $\underset{x_{i}\in\Omega_{i}}{\sup}\left\Vert \mathcal{F}_{i}-\Phi_{i}\left(x_{i},\theta_{i}^{*}\right)\right\Vert <\varepsilon_{i}$
for $i\in\mathcal{I}$.\footnote{It is not known how to obtain a bound $\bar{\theta}_{i}$ on such
parameter $\theta_{i}^{*}$ for an arbitrary $\varepsilon_{i}$, which
causes difficulties in constructing the bounded search space $\Theta_{i}$.
Therefore, we allow $\bar{\theta}_{i}$ to be arbitrarily selected
in the above analysis, at the loss of guarantees on the approximation
accuracy. Although the bound $\varepsilon_{i}$ which bounds $\underset{x_{i}\in\Omega_{i}}{\sup}\left\Vert \mathcal{F}_{i}-\Phi_{i}\left(x_{i},\theta_{i}^{*}\right)\right\Vert $
might no longer be arbitrary in this case, it would still exist due
to the continuity of the function being estimated denoted by index
$i$ and $\Phi_{i}$.} Based on this property, the unknown terms, $M(q),\ C(q,\dot{q}),\ G(q),$
and $F(\dot{q})$ can be modeled as
\begin{align}
\text{\text{vec}}\left(M(q)\right) & =\Phi_{M}(x_{M},\theta_{M}^{*})+\varepsilon_{M}(x_{M}),\label{eq:M_DNN_approx}\\
\text{\text{vec}}\left(C(q,\dot{q})\right) & =\Phi_{C}(x_{C},\theta_{C}^{*})+\varepsilon_{C}(x_{C}),\label{eq:Vm_DNN_approx}\\
G(q) & =\Phi_{G}(x_{G},\theta_{G}^{*})+\varepsilon_{G}(x_{G}),\label{eq:G_DNN_approx}\\
F(\dot{q}) & =\Phi_{F}(x_{F},\theta_{F}^{*})+\varepsilon_{F}(x_{F}),\label{eq:F_DNN_approx}
\end{align}
respectively, where $\Phi_{i}(x_{i},\theta_{i}^{*})\in\mathbb{R}^{L_{out,i}}$
for $i\in\mathcal{I}$ are the DNNs with weights which yield the best
approximation of $\mathcal{F}_{i}$ in the search space $\Theta_{i}$,
$L_{out,i}=n^{2}$ for $i\in\left\{ M,C\right\} $, and $L_{out,i}=n$
for $i\in\left\{ F,G\right\} $. The inputs to the individual DNNs
are defined as $x_{M}\triangleq q$, $x_{C}\triangleq\left[q,\dot{q}\right]^{\top}$,
$x_{G}\triangleq q$, $x_{F}\triangleq\dot{q}$. The unknown function
approximation errors are denoted by $\varepsilon_{M}(x_{M}):\mathbb{R}^{n}\rightarrow\mathbb{R}^{n^{2}}$,
$\varepsilon_{C}(x_{C}):\mathbb{R}^{2n}\rightarrow\mathbb{R}^{n^{2}}$,
$\varepsilon_{G}(x_{G}):\mathbb{R}^{n}\rightarrow\mathbb{R}^{n}$,
and $\varepsilon_{F}(x_{F}):\mathbb{R}^{n}\rightarrow\mathbb{R}^{n}$
for $\Phi{}_{M}(x_{M},\theta_{M}),\ \Phi_{C}(x_{C},\theta_{C}),\ \Phi_{G}(x_{G},\theta_{G}),$
and $\Phi_{F}(x_{F},\theta_{F})$, respectively. The following assumption
is made to facilitate the subsequent control development.
\begin{assum}
\label{assum:varepsilon_prime_bound}Given any compact set $\Omega_{i}$,
there exist known constants $\overline{\varepsilon_{i}},\overline{\varepsilon_{i}^{\prime}}\in\mathbb{R}_{>0}$
such that the function approximation error and it's partial derivative\footnote{See \cite[Thm. 4.1]{Pinkus1999} for an analysis of the function approximation
properties of partial derivatives.} with respect to the input can be bounded as $\varepsilon_{i}(x_{i})\leq\overline{\varepsilon}_{i}$
and $\left\Vert \frac{\partial}{\partial x_{i}}\varepsilon_{i}(x_{i})\right\Vert \leq\overline{\varepsilon_{i}^{\prime}}$
for all $i\in\mathcal{I}$ and $x_{i}\in\Omega_{i}$.
\end{assum}
\begin{rem}
Assumption \ref{assum:varepsilon_prime_bound} is reasonable because
in practice the user can select $\overline{\theta}_{i}$ a priori
and $\overline{\varepsilon}_{i},\overline{\varepsilon_{i}^{\prime}}$
can subsequently be prescribed for $i\in\mathcal{I}$ using a conservative
estimate whose feasibility can be verified using heuristic methods
(e.g., Monte Carlo search). Notably, DNN architectures which contain
spectral normalization layers as in \cite{Shi.Shi.ea2019} inherently
involve bounded weights since the weight matrices are normalized by
their spectral norms.
\end{rem}
Applying the vectorization operator and its properties on $M$ and
$C$, and substituting (\ref{eq:M_DNN_approx})-(\ref{eq:F_DNN_approx})
into (\ref{eq:CLES1}) yields the open loop error system
\begin{align}
M(q)\dot{r} & =\tau-\Phi_{G}(x_{G},\theta_{G}^{*})-\Phi_{F}(x_{F},\theta_{F}^{*})\nonumber \\
 & \quad-\left(\text{\ensuremath{\left(\dot{q}_{d}-\alpha e\right)}}^{\top}\otimes I_{n}\right)\left(\Phi_{C}(x_{C},\theta_{C}^{*})+\varepsilon_{C}(x_{C})\right)\nonumber \\
 & \quad-\left(\text{\ensuremath{\left(\ddot{q}_{d}-\alpha\dot{e}\right)}}^{\top}\otimes I_{n}\right)\bigl(\Phi_{M}(x_{M},\theta_{M}^{*})\nonumber \\
 & \quad+\varepsilon_{M}(x_{M})\bigr)-\varepsilon_{G}(x_{G})-\varepsilon_{F}(x_{F})-C(q,\dot{q})r.\label{eq:CLES2}
\end{align}
Based on the subsequent stability analysis, the adaptive\textbf{ }SS-LbPINN\textbf{
}control input is defined as
\begin{align}
\tau & \triangleq\left(\text{\ensuremath{\left(\dot{q}_{d}-\alpha e\right)}}^{\top}\otimes I_{n}\right)\Phi_{C}(x_{C},\hat{\theta}_{C})+\Phi_{G}(x_{G},\hat{\theta}_{G})\nonumber \\
 & \quad+\Phi_{F}(x_{F},\hat{\theta}_{F})-k_{1}r-e\nonumber \\
 & \quad+\left((\ddot{q}_{d}-\alpha\dot{e})^{\top}\otimes I_{n}\right)\Phi_{M}(x_{M},\hat{\theta}_{M})-\text{sgn\ensuremath{(r)}}\Bigl(k_{2}\nonumber \\
 & \quad+k_{3}\left\Vert (\dot{q}_{d}-\alpha e)^{\top}\otimes I_{n}\right\Vert \left.+k_{4}\left\Vert (\ddot{q}_{d}-\alpha\dot{e}){}^{\top}\otimes I_{n}\right\Vert \right),\label{eq:tau}
\end{align}
where $k_{1},k_{2},k_{3},k_{4}\in\mathbb{R}_{>0}$ are a user-defined
control gains, $\Phi_{i}(x_{i},\hat{\theta}_{i})\in\mathbb{R}^{L_{out,i}}$
denotes the DNN estimate of the unknown matrices, and the individual
adaptive DNN weight estimates are denoted as $\hat{\theta}_{i}\in\mathbb{R}^{\varkappa_{i}}$,
for all $i\in\mathcal{I}$. To address the additional level of complexity
due to the nonlinearity in the DNN, a first order Taylor series approximation-based
error model is given by \cite[Eq. 22]{Patil.Le.ea2022}
\begin{equation}
\Phi_{i}(x_{i},\theta_{i}^{*})-\Phi_{i}(x_{i},\hat{\theta}_{i})=\Phi^{\prime}(x_{i},\hat{\theta}_{i})\tilde{\theta}_{i}+R_{i}(x_{i},\tilde{\theta}_{i}),\label{eq: TaylorSeriesApprox}
\end{equation}
where $R_{i}:\mathbb{R}^{L_{in,i}}\times\mathbb{R}^{\varkappa_{i}}\to\mathbb{R}^{L_{out,i}}$
denotes the Lagrange remainder, $\Phi^{\prime}(x_{i},\hat{\theta}_{i})$
denotes the Jacobian described in Section \ref{subsec:DNNModel},
and $\tilde{\theta}_{i}\triangleq\theta_{i}-\hat{\theta}_{i}$ denotes
the parameter estimation error for $i\in\mathcal{I}$. Using (\ref{eq:CLES2})-(\ref{eq: TaylorSeriesApprox})
the closed loop error system in (\ref{eq:CLES2}) can be written as
\begin{align}
M(q)\dot{r} & =-k_{1}r-e-C(q,\dot{q})r-\text{sgn\ensuremath{(r)}}\Bigl(k_{2}\nonumber \\
 & \quad+k_{3}\left\Vert (\dot{q}_{d}-\alpha e)^{\top}\otimes I_{n}\right\Vert \nonumber \\
 & \quad+\left.k_{4}\left\Vert (\ddot{q}_{d}-\alpha\dot{e}){}^{\top}\otimes I_{n}\right\Vert \right)\nonumber \\
 & \quad-\Bigl(\text{\ensuremath{\left(\ddot{q}_{d}-\alpha\dot{e}\right)}}^{\top}\otimes I_{n}\Bigr)\Bigl(\Phi^{\prime}(x_{M},\hat{\theta}_{M})\tilde{\theta}_{M}+N_{M}\Bigr)\nonumber \\
 & \quad-\Bigl(\text{\ensuremath{\left(\dot{q}_{d}-\alpha e\right)}}^{\top}\otimes I_{n}\Bigr)\Bigl(\Phi^{\prime}(x_{C},\hat{\theta}_{C})\tilde{\theta}_{C}+N_{C}\Bigr)\nonumber \\
 & \quad-\Phi^{\prime}(x_{G},\hat{\theta}_{G})\tilde{\theta}_{G}-N_{G}-\Phi^{\prime}(x_{F},\hat{\theta}_{F})\tilde{\theta}_{F}-N_{F},\label{eq:CLES4}
\end{align}
where the auxiliary function $N_{i}\in\mathbb{R}^{L_{out,i}}$ is
defined as $N_{i}\triangleq\varepsilon_{i}(x_{i})+R_{i}(x_{i},\tilde{\theta}_{i})$
for all $i\in\mathcal{I}$.

\section{Skew-Symmetric Prediction Error and Adaptive Update Laws}

\subsection{Skew-Symmetric Prediction Error Formulation\label{subsec:Skew-Symmetric-Prediction-Error}}

To incorporate known properties of the system, a skew-symmetric prediction
error is developed to inform the update law. Specifically, the skew-symmetry
relation stated in Property \ref{Prop:SkewSymmetric} describes an
inherent structure in the system dynamics and is used to develop a
penalty in the adaption. The developed skew-symmetric prediction error
penalizes updates that violate the skew-symmetry constraint. To quantify
the skew-symmetric objective, a skew-symmetric prediction error $\widetilde{E}\in\mathbb{R}$
is defined as
\begin{align}
\widetilde{E}(t) & \triangleq E-\widehat{E},\label{eq:PredictionError}
\end{align}
where $\widehat{E}$ is the subsequently designed skew symmetric estimate,
$E\triangleq\int_{0}^{t}\xi^{\top}\left(\frac{d}{d\varsigma}M(x_{M}(\varsigma))-2C(x_{C}(\varsigma))\right)\xi d\varsigma$,
and $\xi\in\mathbb{R}^{n}$ denotes a vector of positive control gains.
Using Property \ref{Prop:SkewSymmetric}, and the fact that $E=0$,
indicates that $\widetilde{E}(t)=-\widehat{E}$. Quantitatively, the
objective of the skew-symmetric constraint is to ensure $\left\Vert \widetilde{E}(t)\right\Vert \to0$
as $t\to\infty$. The corresponding skew-symmetric estimate given
by $\widehat{E}\in\mathbb{R}$ is defined as a Filippov solution of%
\begin{comment}
The integral in the equation below is motivated by the fact that we
will need $\dot{\widehat{E}}$ in the Lyapunov-analysis, because we're
already using $\frac{d}{dt}\Phi(x_{M},\hat{\theta}_{M})$ we'll define
the prediction error as the integral to avoid taking the derivative
a second time
\end{comment}
\begin{align}
\dot{\widehat{E}} & \triangleq\left(\xi^{\top}\otimes\xi^{\top}\right)\biggl(\frac{d}{dt}\Bigl(\Phi(x_{M},\hat{\theta}_{M})\Bigr)-2\Phi(x_{C},\hat{\theta}_{C})+\mu\biggr),\label{eq:SkewSymmetricEstimation}
\end{align}
where $\mu(t)\in\mathbb{R}^{n}$ is given by
\begin{equation}
\mu(t)\triangleq-\widehat{\Phi}_{M}^{\prime}\dot{\hat{\theta}}_{M}+\gamma_{1}\widetilde{E}+\left(\gamma_{2}+\gamma_{3}+\gamma_{4}\left\Vert \dot{x}_{M}\right\Vert \right)\text{sgn}(\widetilde{E}),\label{eq:SkewSymmetricRISETerm}
\end{equation}
where $\gamma_{1},\gamma_{2},\gamma_{3},\gamma_{4}\in\mathbb{R}$
denote constant positive control gains.

To facilitate the subsequent stability analysis, it is desirable to
obtain an analytical expression for $\widetilde{E}$ which allows
the skew-symmetric prediction error to be explicitly related to the
parameter estimation error of the skew-symmetric matrix estimates
(i.e., $\tilde{\theta}_{M}$ and $\tilde{\theta}_{C}$). Therefore,
an equivalent representation for $E$ is developed using properties
of vectorization and the Kronecker product which allows $E$ to be
expressed as $E=\int_{0}^{t}\left(\xi^{\top}\otimes\xi^{\top}\right)\left(\text{vec}\left(\frac{d}{d\varsigma}M(x_{M}(\varsigma))\right)-\text{vec}(2C(x_{C}(\varsigma)))\right)d\varsigma.$
Using (\ref{eq:Vm_DNN_approx}) and the time derivative of (\ref{eq:M_DNN_approx}),
and the fact that $E$ is differentiable almost everywhere, then its
time-derivative, where it exists, is given by
\begin{align}
\dot{E} & =\left(\xi^{\top}\otimes\xi^{\top}\right)\Bigl(\frac{d}{dt}\left(\Phi(x_{M},\theta_{M}^{*})+\varepsilon(x_{M})\right)\nonumber \\
 & \quad-2\left(\Phi(x_{C},\theta_{C}^{*})+\varepsilon(x_{C})\right)\Bigr).\label{eq:AnalyticalE}
\end{align}
Substituting (\ref{eq:SkewSymmetricEstimation}) and (\ref{eq:AnalyticalE})
into (\ref{eq:PredictionError}) and applying (\ref{eq: TaylorSeriesApprox})\textbf{
}yields the skew symmetric prediction error $\widetilde{E}$ given
by the Filippov solution of
\begin{align}
\dot{\widetilde{E}} & =\left(\xi^{\top}\otimes\xi^{\top}\right)\Bigl(\frac{d}{dt}\left(\Phi^{\prime}(x_{M},\hat{\theta}_{M})\tilde{\theta}_{M}+N_{M}\right)\nonumber \\
 & \quad-2\left(\Phi^{\prime}(x_{C},\hat{\theta}_{C})\tilde{\theta}_{C}+N_{C}\right)-\mu\Bigr).\label{eq:AnalyticalPredError1}
\end{align}
Applying the chain rule to $\frac{d}{dt}(\Phi^{\prime}(x_{M},\hat{\theta}_{M})\tilde{\theta}_{M})$,
using (\ref{eq:SkewSymmetricRISETerm}), and canceling cross terms,
$\dot{\widetilde{E}}$ can be equivalently expressed as
\begin{align}
\dot{\widetilde{E}} & =\left(\xi^{\top}\otimes\xi^{\top}\right)\Bigl(\frac{d}{dt}\left(\Phi^{\prime}(x_{M},\hat{\theta}_{M})\right)\tilde{\theta}_{M}+\frac{d}{dt}\left(N_{M}\right)\nonumber \\
 & -\gamma_{1}\widetilde{E}-2\left(\Phi^{\prime}(x_{C},\hat{\theta}_{C})\tilde{\theta}_{C}+N_{C}\right)\nonumber \\
 & -\left(\gamma_{2}+\gamma_{3}+\gamma_{4}\left\Vert \dot{x}_{M}\right\Vert \right)\text{sgn}(\widetilde{E})\Bigr).\label{eq:dot_tilde_E}
\end{align}

\subsection{SS-LbPINN Weight Adaptation Laws}

Lyapunov-based adaptation laws which incorporate skew-symmetric penalties
enable continuous, real-time parameter updates which are guided by
the performance of the system and the underlying known physical properties.
Specifically, the subsequent adaptation laws for the inertial and
centripetal-Coriolis estimates incorporate the skew-symmetric prediction
error term developed in Section \ref{subsec:Skew-Symmetric-Prediction-Error}.
The Lyapunov-based adaptation laws for the SS-LbPINN are defined as
\begin{align}
\dot{\hat{\theta}}_{M} & \triangleq\text{proj}\Bigl(\Gamma_{M}\Bigl(\frac{d}{dt}\left(\Phi^{\prime}\left(x_{M},\hat{\theta}_{M}\right)\right)^{\top}\left(\xi\otimes\xi\right)\widetilde{E}\nonumber \\
 & \quad-\Phi^{\prime\top}\left(x_{M},\hat{\theta}_{M}\right)\left((\ddot{q}_{d}-\alpha\dot{e})^{\top}\otimes I_{n}\right)^{\top}r\Bigr),\Theta_{M}\Bigr),\label{eq:M_hat_dot}\\
\dot{\hat{\theta}}_{C} & \triangleq\text{proj}\Bigl(\Gamma_{C}\Bigl(-2\Phi^{\prime\top}(x_{C},\hat{\theta}_{C})\left(\xi\otimes\xi\right)\widetilde{E}\nonumber \\
 & \quad-\Phi^{\prime\top}(x_{C},\hat{\theta}_{C})\left((\dot{q}_{d}-\alpha e)^{\top}\otimes I_{n}\right)^{\top}r\Bigr),\Theta_{C}\Bigr),\label{eq:V_hat_dot}\\
\dot{\hat{\theta}}_{F} & \triangleq\text{proj}\Bigl(-\Gamma_{F}\Phi^{\prime\top}(x_{F},\hat{\theta}_{F})r,\Theta_{F}\Bigr),\label{eq:F_hat_dot}\\
\dot{\hat{\theta}}_{G} & \triangleq\text{proj}\Bigl(-\Gamma_{G}\Phi^{\prime\top}(x_{G},\hat{\theta}_{G})r,\Theta_{G}\Bigr),\label{eq:G_hat_dot}
\end{align}
where $\Gamma_{i}\in\mathbb{R}^{\varkappa_{i}\times\varkappa_{i}}$
for $i\in\mathcal{I}$ is a positive definite user-defined adaptation
gain matrix and $\text{proj}\left(\cdot\right)$ denotes a continuous
projection operator (cf. \cite[Appendix E]{Krstic.Kanellakopoulos.ea1995})
which ensures $\hat{\theta}_{i}(t)\in\mathbb{R}^{\varkappa_{i}}\triangleq\left\{ \theta_{i}\in\mathbb{R}^{\varkappa_{i}}:\left\Vert \theta_{i}\right\Vert \leq\overline{\theta}_{i}\right\} $
for all $t\in\mathbb{R}_{\geq0}$ and $i\in\mathcal{I}$.

\section{Stability Analysis}

Let $z\triangleq\left[e^{\top},r^{\top},\widetilde{E}\right]^{\top}\in\mathbb{R}^{2n+1}$
denote the concatenated state. Using \cite[Thm. 1]{Patil.Fallin.ea2025}
and the user-selected bound on the search space, the\textbf{ }Lagrange
remainder can be bounded as \textbf{$\left\Vert R_{i}(x_{i},\tilde{\theta}_{i})\right\Vert \leq4\rho_{0,i}\left(\left\Vert x_{i}\right\Vert \right)\overline{\theta}_{i}^{2}$
}where $\rho_{0,i}:\mathbb{R}_{\geq0}\to\mathbb{R}_{\geq0}$ is a
non-decreasing positive function of the form $\rho_{0,i}(\left\Vert x_{i}\right\Vert )=a_{2,i}\left\Vert x_{i}\right\Vert ^{2}+a_{1,i}\left\Vert x_{i}\right\Vert +a_{0,i}$
with some constants $a_{2,i},a_{1,i},a_{0,i}\in\mathbb{R}_{>0}$,
for all $i\in\mathcal{I}$. Using the same process as \cite{Patil.Fallin.ea2025},
applying the chain rule to find the time derivative, and using the
activation function bounds in (\ref{eq:Activation Bounds}), the time
derivative of $R_{M}(x_{M},\tilde{\theta}_{M})$ can be bounded as
$\left\Vert \frac{d}{dt}R_{M}(x_{M},\tilde{\theta}_{M})\right\Vert \leq2\overline{\theta}_{M}^{2}\rho_{0,M}(\left\Vert x_{M}\right\Vert )+4\overline{\theta}_{M}^{2}\rho_{T}(\left\Vert x_{M}\right\Vert ,\left\Vert \dot{x}_{M}\right\Vert )$
where $\rho_{T}:\mathbb{R}_{\geq0}\to\mathbb{R}_{\geq0}$ is a non-decreasing
positive function of the form $\rho_{T}(\left\Vert x_{M}\right\Vert ,\left\Vert \dot{x}_{M}\right\Vert )=b_{h,i}\left\Vert x_{i}\right\Vert ^{h}+b_{h,i}\left\Vert \dot{x}_{i}\right\Vert ^{h}+b_{h-1,i}\left\Vert x_{i}\right\Vert ^{h-1}+b_{h-1,i}\left\Vert \dot{x}_{i}\right\Vert ^{h-1}+\dots+b_{0,i}$
for some $b_{h,i},\dots,b_{0,i}\in\mathbb{R}_{>0}$. The terms $N_{i}$
and $\frac{d}{dt}N_{M}$ can be bounded as $\left\Vert N_{i}\right\Vert \leq\left\Vert R_{i}(x_{i},\tilde{\theta}_{i})+\varepsilon_{i}(x_{i})\right\Vert $
for $i\in\mathcal{I}$ and $\left\Vert \frac{d}{dt}N_{M}\right\Vert \leq\left\Vert \frac{d}{dt}R_{M}(x_{M},\tilde{\theta}_{M})\right\Vert +\left\Vert \frac{d}{dt}\varepsilon_{M}(x_{M})\right\Vert $,
respectively, where $\frac{d}{dt}\varepsilon_{M}(x_{M})=\frac{\partial\varepsilon(x_{M})}{\partial x_{M}}\dot{x}_{M}$.
The inputs $x_{i}$ can be upper-bounded as $\left\Vert x_{i}\right\Vert \leq\overline{z_{i}}+\overline{q}_{d,i}$
where, $\overline{z_{M}},\overline{z_{G}}\triangleq\left\Vert z\right\Vert $,
$\overline{z_{C}}\triangleq(\alpha+2)\left\Vert z\right\Vert $, $\overline{z_{F}}\triangleq(\alpha+1)\left\Vert z\right\Vert $,
$\overline{q_{d,M}},\overline{q_{d,G}}\triangleq\overline{q}_{d},$
$\overline{q_{d,C}}\triangleq\overline{q}_{d}+\overline{\dot{q}_{d}}$,
and $\overline{q_{d,F}}\triangleq\overline{\dot{q}_{d}}$. The time
derivative of the input $x_{M}$ can be upper bounded as $\left\Vert \dot{x}_{M}\right\Vert \leq\overline{z}_{dt,M}+\overline{q}_{dt,M}$
where $\overline{z_{dt,M}}\triangleq(\alpha+1)\left\Vert z\right\Vert $
and $\overline{q_{dt,M}}\triangleq\overline{\dot{q}_{d}}$. Then,
$\left\Vert N_{i}\right\Vert \leq4\overline{\theta}_{i}^{2}\rho_{0,i}(\overline{z}_{i}+\overline{q}_{d,i})+\overline{\varepsilon}_{i}$,
and hence, from Assumption \ref{assum:varepsilon_prime_bound}, $\left\Vert \frac{d}{dt}N_{M}\right\Vert \leq2\overline{\theta}_{M}^{2}\rho_{0,M}(\overline{z}_{M}+\overline{q}_{d,M})+4\overline{\theta}_{M}^{2}\rho_{T}(\overline{z}_{M}+\overline{q}_{d,M}+\overline{z}_{dt,M}+\overline{q}_{dt,M})+\overline{\varepsilon_{M}^{\prime}}\left\Vert \dot{x}_{M}\right\Vert $,
which can be further bounded as
\begin{equation}
\left\Vert N_{i}\right\Vert \leq4\overline{\theta}_{i}^{2}\left(\rho_{1,i}(\overline{z_{i}})+\rho_{2,i}(\overline{q_{d,i}})\right)+\overline{\varepsilon}_{i}\label{eq:N_i_bound1}
\end{equation}
 and
\begin{align}
\left\Vert \frac{d}{dt}N_{M}\right\Vert  & \leq2\overline{\theta}_{M}^{2}\left(\rho_{1,M}(\overline{z}_{M})+\rho_{2,M}(\overline{q_{d,M}})\right)\nonumber \\
 & \quad+\overline{\varepsilon_{M}^{\prime}}\left\Vert \dot{x}_{M}\right\Vert +4\overline{\theta}_{M}^{2}\rho_{T1}(\overline{z_{M}}+\overline{z}_{dt,M})\nonumber \\
 & \quad+4\overline{\theta}_{M}^{2}\rho_{T2}\left(\overline{q}_{d,M}+\overline{q}_{dt,M}\right)\label{eq:d_dt_N_M_Bound1}
\end{align}
where $\rho_{1,i}:\mathbb{R}_{\geq0}\to\mathbb{R}_{\geq0}$, $\rho_{2,i}:\mathbb{R}_{\geq0}\to\mathbb{R}_{\geq0}$,
for all $i\in\mathcal{I}$, $\rho_{T1}:\mathbb{R}_{\geq0}\to\mathbb{R}_{\geq0}$,
and $\rho_{T2}:\mathbb{R}_{\geq0}\to\mathbb{R}_{\geq0}$ are non-decreasing
positive functions. Let $\overline{\rho_{1,i}}(\left\Vert z\right\Vert )\triangleq4\overline{\theta}_{i}^{2}\left(\rho_{1,i}(\overline{z_{i}})-\rho_{1,i}(0)\right)$
for $i\in\mathcal{I}$ and $\overline{\rho_{T1}}(\left\Vert z\right\Vert )=4\overline{\theta}_{M}\left(\rho_{T1}\left((\alpha_{1}+2)\left\Vert z\right\Vert \right)-\rho_{T1}(0)\right)$.
Multiplying, $\overline{\rho_{1,i}}(\left\Vert z\right\Vert )$ and
$\overline{\rho_{T1}}(\left\Vert z\right\Vert )$ by $\frac{\sqrt{c_{s,i}}}{\sqrt{c_{s,i}}}$
and $\frac{\sqrt{c_{sT}}}{\sqrt{c_{sT}}}$, respectively, where $c_{s,i},c_{sT}\in\mathbb{R}_{>0}$
are user-defined constants, and applying the triangle-inequality,
$\overline{\rho_{1,i}}$ and $\overline{\rho_{T1}}$ can be bounded
as $\overline{\rho_{1,i}}(\left\Vert z\right\Vert )\leq\frac{\overline{\rho_{1,i}}^{2}(\left\Vert z\right\Vert )}{2c_{s,i}}+\frac{c_{s,i}}{2}$
and $\overline{\rho_{T1}}(\left\Vert z\right\Vert )\leq\frac{\overline{\rho_{T1}}^{2}(\left\Vert z\right\Vert )}{2c_{sT}}+\frac{c_{sT}}{2}$
where there exists $\rho_{i}$ and $\rho_{dt}$ such that $\frac{\overline{\rho_{1,i}}^{2}(\left\Vert z\right\Vert )}{2c_{s,i}}\leq\rho_{i}(\left\Vert z\right\Vert )\left\Vert z\right\Vert $
for $i\in\mathcal{I}$ and $\frac{\overline{\rho_{T1}}^{2}(\left\Vert z\right\Vert )}{2c_{sT}}\leq\rho_{dt}(\left\Vert z\right\Vert )\left\Vert z\right\Vert $.
Then, using (\ref{eq:N_i_bound1}), the developed bounds, and adding
and subtracting $4\overline{\theta}_{i}^{2}\rho_{1,i}(0)$, then $N_{i}$
can be bounded as
\begin{align}
\left\Vert N_{i}\right\Vert  & \leq\rho_{i}(\left\Vert z\right\Vert )\left\Vert z\right\Vert +4\overline{\theta}_{i}^{2}\left(\rho_{2,i}(\overline{q}_{i})+\rho_{1,i}(0)\right)\nonumber \\
 & \quad+\overline{\varepsilon_{i}}+\frac{1}{2}c_{s,i},\label{eq:N_i_bound_final}
\end{align}
 for $i\in\mathcal{I}$. Similarly, using (\ref{eq:N_i_bound1}) and
(\ref{eq:d_dt_N_M_Bound1}),
\begin{align}
\left\Vert \frac{d}{dt}N_{M}\right\Vert  & \leq\left(\rho(\left\Vert z\right\Vert )+\rho_{dt}(\left\Vert z\right\Vert )\right)\left\Vert z\right\Vert +2\overline{\theta}_{M}^{2}\rho_{2,M}\left(\overline{q}_{d}\right)\nonumber \\
 & \quad+4\overline{\theta}_{M}^{2}\rho_{T2}(\overline{q_{d}}+\overline{\dot{q}_{d}})+\overline{\varepsilon_{M}^{\prime}}\left\Vert \dot{x}_{M}\right\Vert +\frac{c_{sT}}{2}\nonumber \\
 & \quad+2\overline{\theta}_{M}^{2}\rho_{1,M}(0)+4\overline{\theta}_{M}^{2}\rho_{T1}(0)+\frac{c_{s}}{2}.\label{eq:d_dt_N_M_bound_final}
\end{align}

To facilitate the subsequent stability analysis, let the concatenated
state vector $\zeta\in\mathbb{R}^{\psi}$ be defined as $\zeta\triangleq[e^{\top},r^{\top},\widetilde{E},\sum_{i\in\mathcal{I}}\tilde{\theta}_{i}^{\top}]^{\top}$where
$\psi\triangleq2n+1+\sum_{i\in\mathcal{I}}\varkappa_{i}$. Since the
approximation capabilities of DNNs stated in (\ref{eq:M_DNN_approx})-(\ref{eq:F_DNN_approx})
hold only on the compact domains $\Omega_{i}$, for $i\in\mathcal{I}$,
the subsequent stability analysis requires ensuring $x_{i}(t)\in\Omega_{i}$
for all $t\in[t_{0},\infty)$. This is achieved by demonstrating that
$\zeta$ is constrained to a compact domain. To that end, consider
the compact domain $\mathcal{D}\triangleq\left\{ \sigma\in\mathbb{R}^{\psi}:\left\Vert \sigma\right\Vert <\chi\right\} $
in which $\zeta$ is supposed to lie. It follows that if $\left\Vert \zeta\right\Vert <\chi$,
then $\left\Vert x_{i}\right\Vert <\overline{\chi}_{i}+\overline{q}_{d,i}$
for $i\in\mathcal{I}$ where $\overline{\chi_{M}},\overline{\chi_{G}}\triangleq\chi$,
$\overline{\chi_{C}}\triangleq(\alpha+2)\chi$, and $\overline{\chi_{F}}\triangleq(\alpha+1)\chi$.
Therefore, if $\Omega_{i}\triangleq\left\{ \sigma\in\mathbb{R}^{L_{in,i}}:\left\Vert \sigma\right\Vert <\overline{\chi_{i}}+\overline{q}_{d,i}\right\} $
for $i\in\mathcal{I}$, then $\zeta\in\mathcal{D}$ implies $x_{i}\in\Omega_{i}$
for all $i\in\mathcal{I}$. Let $\lambda_{1}\triangleq\text{min}\left\{ \alpha,k_{1},\gamma_{1}\left(\xi^{\top}\otimes\xi^{\top}\right)\right\} $,
$\lambda_{2}\in\mathbb{R}_{>0}$ be the desired convergence rate,
$\rho_{3}\triangleq\rho_{2,i}(\overline{q}_{i})+\rho_{1,i}(0)$ for
$i\in\mathcal{I}$, $P_{1}\left(\left\Vert z\right\Vert \right)=\sum_{i\in\mathcal{I}}\overline{Q_{i}}\rho_{i}(\left\Vert z\right\Vert )+\left\Vert \xi^{\top}\otimes\xi^{\top}\right\Vert \left(2\rho_{V}(\left\Vert z\right\Vert )+\rho_{M}(\left\Vert z\right\Vert )+\rho_{dt}(\left\Vert z\right\Vert )\right)$
where $\overline{Q_{M}}\triangleq\left\Vert (\ddot{q}_{d}-\alpha\dot{e}){}^{\top}\otimes I_{n}\right\Vert $,
$\overline{Q_{V}}=\left\Vert \text{\ensuremath{\left(\dot{q}_{d}-\alpha e\right)}}^{\top}\otimes I_{n}\right\Vert $,
and $\overline{Q_{F}},\overline{Q_{G}}=1$, and $P_{1}\left(\left\Vert z\right\Vert \right)\leq P\left(\left\Vert z\right\Vert \right)$
where $P$ is a positive strictly increasing and invertible function.
Let the set of stabilizing initial conditions $\mathcal{S}\subset\mathcal{D}$
be defined as
\begin{equation}
\mathcal{S}\triangleq\left\{ \sigma\in\mathbb{R}^{\psi}:\left\Vert \sigma(t_{0})\right\Vert <\sqrt{\frac{\beta_{1}}{\beta_{2}}}\chi\right\} ,\label{eq:InitialConditions}
\end{equation}
where $\beta_{1}\triangleq\text{min}\left\{ \frac{1}{2},\frac{1}{2}m_{1},\frac{1}{2},\frac{1}{2}\underset{i\in\mathcal{I}}{\text{min}}(\lambda_{\text{min}}(\Gamma_{i}))\right\} $,
$\beta_{2}=\text{max}\left\{ \frac{1}{2},\frac{1}{2}m_{2},\frac{1}{2},\frac{1}{2}\underset{i\in\mathcal{I}}{\text{max}}(\lambda_{\text{max}}(\Gamma_{i}))\right\} $\footnote{Recall, $m_{1}$and $m_{2}$ denote known constants defined in Property
\ref{Prop:PDInertia}.}. The following analysis indicates that convergence starting from
an arbitrary $\zeta(0$) is ensured by selecting
\[
\chi=P^{-1}\left(\lambda_{1}-\lambda_{2}\right),
\]
with a sufficiently large $\lambda_{1}-\lambda_{2}$. The following
theorem establishes the convergence of the tracking error and skew-symmetric
prediction error using the developed SS-LbPINN adaptive controller,
update laws, and prediction error formulation. 
\begin{thm}
Consider the Filippov regularization of the dynamical system in (\ref{eq:dynamics}),
the controller given in (\ref{eq:tau}), and the adaptation laws in
(\ref{eq:M_hat_dot})-(\ref{eq:G_hat_dot}). Provided Assumptions
\ref{assm:activation bounds}-\ref{assum:varepsilon_prime_bound}\textbf{
}hold,\textbf{ $\zeta(t_{0})\in\mathcal{\mathcal{S}}$},\textbf{ }and
the following gain conditions are satisfied, $\lambda_{1}>\lambda_{2}+P\left(\sqrt{\frac{\beta_{2}}{\beta_{1}}\left\Vert z(t_{0})\right\Vert ^{2}}\right)$,
$k_{2}>4\sum_{j=\{F,G\}}\overline{\theta}_{j}^{2}\rho_{3,j}+\overline{\varepsilon_{j}}+\frac{1}{2}c_{s,j}$,
$k_{3}>4\overline{\theta}_{V}^{2}\rho_{3,V}+\overline{\varepsilon_{V}}+\frac{1}{2}c_{s,V}$,
$k_{4}>4\overline{\theta}_{M}^{2}\rho_{3,M}+\overline{\varepsilon_{M}}+\frac{1}{2}c_{s,M}$,
$\gamma_{2}>2\overline{\theta}_{M}^{2}\rho_{3,M}+4\overline{\theta}_{M}^{2}\rho_{T2}(\overline{q_{d}}+\overline{\dot{q}_{d}})+4\overline{\theta}_{M}^{2}\rho_{T1}(0)+\frac{c_{sT}}{2}+\frac{c_{s}}{2}$,
$\gamma_{3}>8\overline{\theta}_{V}^{2}\rho_{3,V}+\overline{\varepsilon_{V}}+\frac{1}{2}c_{s,V}$,
and $\gamma_{4}>\overline{\varepsilon_{M}^{\prime}}$ , the tracking
and skew-symmetric prediction errors asymptotically converge to zero.
Specifically, $\left\Vert e(t)\right\Vert \to0,$ $\left\Vert r(t)\right\Vert \to0$,
and $\left\Vert \widetilde{E}(t)\right\Vert \to0$ as $t\to\infty$.
\end{thm}
\begin{pf}
Consider the candidate Lyapunov function $\mathcal{V}_{L}\in\mathbb{R}_{\geq0}$
which is a Lipschitz continuous positive definite function defined
as
\begin{equation}
\mathcal{V}_{L}(\zeta,t)\triangleq\frac{1}{2}e^{\top}e+\frac{1}{2}r^{\top}Mr^{\top}+\frac{1}{2}\widetilde{E}^{2}+\frac{1}{2}\sum_{i\in\mathcal{I}}\tilde{\theta}_{i}^{\top}\Gamma_{i}^{-1}\tilde{\theta}_{i}.\label{eq:Lyapunov}
\end{equation}
The candidate Lyapunov function in (\ref{eq:Lyapunov}) satisfies
the inequality
\begin{equation}
\beta_{1}\left\Vert \zeta\right\Vert ^{2}\leq\mathcal{V}_{L}(\zeta,t)\leq\beta_{2}\left\Vert \zeta\right\Vert ^{2}.\label{eq:beta<V<beta}
\end{equation}
Let $\dot{\zeta}=\mathcal{H}(\zeta,t)$, where $\mathcal{H}(\zeta,t)\in\mathbb{R}_{\geq0}\times\mathbb{R}^{\psi}$
denotes the right-hand side of the closed-loop error signals. Using
Filippov's theory of differential inclusions \cite{Filippov1964,Filippov1988,Smirnov2002,Aubin2008},
the existence of solutions can be established for $\dot{\zeta}\in\text{K}\left[\mathcal{H}(\zeta,t)\right](\zeta)$,
where $\text{K}[\mathcal{H}(\zeta,t)]\triangleq\underset{\delta>0}{\cap}\underset{\mu L=0}{\cap}\overline{co}\mathcal{H}(B(\zeta,\delta)\setminus L,t),$
the intersection over all sets $L$ of Lebesgue measure zeros is denoted
$\underset{\mu L=0}{\cap}$, $\overline{co}$ denotes convex closure,
and $B(\zeta,\delta)=\left\{ w\in R^{\psi}|\left\Vert \zeta-w\right\Vert <\delta\right\} .$
The time derivative of (\ref{eq:Lyapunov}) exists along the trajectories
of $\dot{\zeta}=\text{K}[\mathcal{H}(\zeta,t)]$ almost everywhere
$(\text{a.e.})$, i.e., for almost all $t\in\left[t_{0},t_{f}\right],$
and $\dot{\mathcal{V}}_{L}\overset{\text{a.e.}}{\in}\dot{\widetilde{\mathcal{V}}}_{L}$
where $\dot{\widetilde{\mathcal{V}}}=\nabla\mathcal{V}_{L}^{\top}$K$\left[\mathcal{H}(\zeta,t)\right]^{\top}$.
Applying Property \ref{Prop:SkewSymmetric} on (\ref{eq:CLES4}),
using (\ref{eq:M_hat_dot})-(\ref{eq:G_hat_dot}), and canceling cross
terms yields
\begin{align}
\dot{\widetilde{\mathcal{V}}}_{L}(\zeta,t) & \subset e^{\top}\left(-\alpha e\right)+r^{\top}\Bigl(-k_{1}r-\text{K}\left[\text{sgn\ensuremath{(r)}}\right]\Bigl(k_{2}\nonumber \\
 & \quad+k_{3}\left\Vert (\dot{q}_{d}-\alpha e)^{\top}\otimes I_{n}\right\Vert \nonumber \\
 & \quad\left.+k_{4}\left\Vert (\ddot{q}_{d}-\alpha\dot{e}){}^{\top}\otimes I_{n}\right\Vert \right)-N_{G}-N_{F}\nonumber \\
 & \quad-\left(\text{\ensuremath{\left(\ddot{q}_{d}-\alpha\dot{e}\right)}}^{\top}\otimes I_{n}\right)N_{M}\nonumber \\
 & \quad-\left(\text{\ensuremath{\left(\dot{q}_{d}-\alpha e\right)}}^{\top}\otimes I_{n}\right)N_{C}\Bigr)\nonumber \\
 & \quad+\widetilde{E}\dot{\widetilde{E}}-\tilde{\theta}_{M}^{\top}\frac{d}{dt}\left(\widehat{\Phi}^{\prime}(x_{M},\hat{\theta}_{M})\right)^{\top}\widetilde{E}\nonumber \\
 & \quad+2\tilde{\theta}_{C}^{\top}\Phi^{\prime\top}(x_{C},\hat{\theta}_{C})\widetilde{E}.\label{eq:V_dot1}
\end{align}
Injecting (\ref{eq:dot_tilde_E}) into (\ref{eq:V_dot1}) and canceling
cross terms yields
\begin{align}
\dot{\mathcal{V}}_{L}(\zeta,t) & \overset{\text{a.e}}{\in}e^{\top}\left(-\alpha e\right)+r^{\top}\Bigl(-k_{1}r-\text{K}\left[\text{sgn\ensuremath{(r)}}\right]\Bigl(k_{2}\nonumber \\
 & \quad+k_{3}\left\Vert (\dot{q}_{d}-\alpha e)^{\top}\otimes I_{n}\right\Vert \nonumber \\
 & \quad\left.+k_{4}\left\Vert (\ddot{q}_{d}-\alpha\dot{e}){}^{\top}\otimes I_{n}\right\Vert \right)-N_{G}-N_{F}\nonumber \\
 & \quad-\left(\text{\ensuremath{\left(\ddot{q}_{d}-\alpha\dot{e}\right)}}^{\top}\otimes I_{n}\right)N_{M}\nonumber \\
 & \quad-\left(\text{\ensuremath{\left(\dot{q}_{d}-\alpha e\right)}}^{\top}\otimes I_{n}\right)N_{C}\Bigr)\nonumber \\
 & \quad+\widetilde{E}\left(\xi^{\top}\otimes\xi^{\top}\right)\Bigl(\frac{d}{dt}\left(N_{M}\right)-2N_{C}-\gamma_{1}\widetilde{E}\nonumber \\
 & \quad-\left(\gamma_{2}+\gamma_{3}+\gamma_{4}\left\Vert \dot{x}_{M}\right\Vert \right)\text{K}\left[\text{sgn}(\widetilde{E})\right]\Bigr).\label{eq:V_dot3}
\end{align}
Applying the bounds developed in (\ref{eq:N_i_bound_final}) and (\ref{eq:d_dt_N_M_bound_final}),
and provided the stated gain conditions are satisfied,\textbf{ }(\ref{eq:V_dot3})
can be bounded as
\begin{align}
\dot{\mathcal{V}}_{L}(\zeta,t) & \overset{\text{a.e.}}{\leq}-\left(\lambda_{1}-P\left(\left\Vert z\right\Vert \right)\right)\left\Vert z\right\Vert ^{2},\label{eq:V_dot4}\\
 & \overset{\text{a.e.}}{\leq}-\lambda_{2}\left\Vert z\right\Vert ^{2},\label{eq:Final_V}
\end{align}
when $\zeta\in\mathcal{D}$. Therefore, when $\zeta$ is initialized
such that $\zeta(t_{0})\in\mathcal{S}$, then using (\ref{eq:Lyapunov})
and (\ref{eq:Final_V}),  the extension of the LaSalle-Yoshizawa corollary
in \cite[Corollary 1]{Fischer.Kamalapurkar.ea2013}\textbf{ }can be
invoked to show that $\lambda_{2}\left\Vert z\right\Vert ^{2}\to0$
as $t\to\infty$, for all $\zeta(t_{0})\in\mathcal{S}$. Based on
the definition of $z$, it follows that $\left\Vert e(t)\right\Vert \to0$,
$\left\Vert r(t)\right\Vert \rightarrow0$, and $\left\Vert \widetilde{E}(t)\right\Vert \to0$
for all $\zeta(t_{0})\in\mathcal{S}$. 

The fact that $\mathcal{V}_{L}(\zeta,t)$ is nonincreasing implies
$\left\Vert \zeta(t)\right\Vert \leq\sqrt{\frac{\mathcal{V}_{L}(\zeta,t)}{\beta_{1}}}\leq\text{\ensuremath{\sqrt{\frac{\mathcal{V}_{L}(\zeta,t_{0})}{\beta_{1}}}}}$
and the fact that $\zeta(t_{0})\in\mathcal{S}$ guarantees that $\zeta(t)\in\mathcal{D}$
for all $t\in[0,\infty)$ and thus $x_{i}\in\Omega_{i}$ for all $t\in[0,\infty)$
and $i\in\mathcal{I}.$ Using (\ref{eq:Lyapunov}) and (\ref{eq:Final_V})
implies $e,r,\widetilde{E},\tilde{\theta}_{M},\tilde{\theta}_{C},\tilde{\theta}_{F},\tilde{\theta}_{G}\in\mathcal{L}_{\infty}.$
Using the fact that $\widetilde{E}\in\mathcal{L}_{\infty}$, the definition
in (\ref{eq:PredictionError}), and the fact that $E=0$, implies
$\widehat{E}\in\mathcal{L}_{\infty}$. Due to the use of the projection
operator, $\hat{\theta}_{i}\in\mathcal{L}_{\infty}$ for all $i\in\mathcal{I}$.
The fact that $q_{d},\dot{q}_{d},e,r\in\mathcal{L}_{\infty}$ implies
$q,\dot{q}\in\mathcal{L}_{\infty}$. Using (\ref{eq:tau}), the fact
that $e,r,q_{d},\dot{q}_{d},\ddot{q}_{d},\,\text{and }\hat{\theta}_{i}\in\mathcal{L}_{\infty}$
implies $x_{i}\in\mathcal{L}_{\infty}$ and thus $\Phi(x_{i},\hat{\theta}_{i})$
is bounded for $i\in\mathcal{I}$ and therefore $\tau$ is bounded. 
\end{pf}

\section{Simulations}

\begin{figure}[h]
\includegraphics[width=1.05\columnwidth]{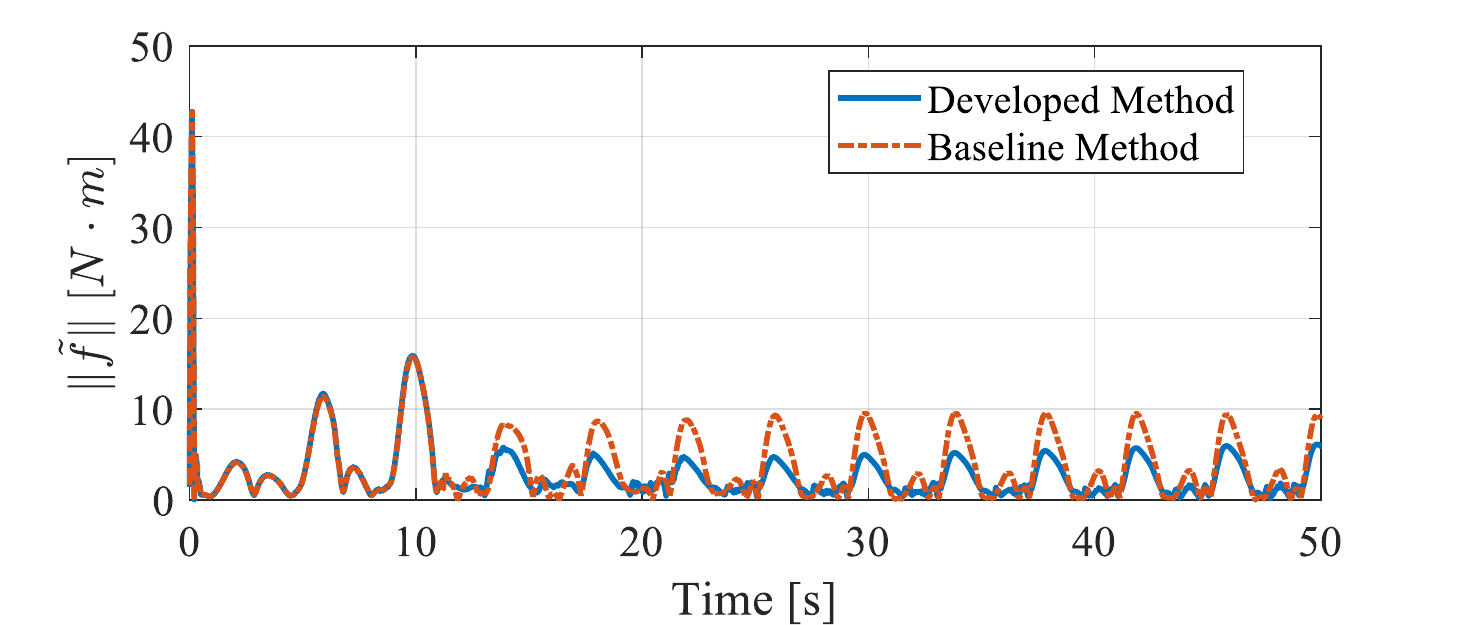}\centering\caption{\label{fig:OveralllFuncApprox}Overall function approximation error
over time for the developed method and the baseline method in \cite{Hart.patil.ea2023}.}
\end{figure}

\begin{table*}
\caption{\label{tab:table}Comparison of Performance (RMS)}

\centering{}%
\begin{tabular}{cccccc}
RMS Values & $\left\Vert \tilde{f}\right\Vert $ & $\left\Vert M-\widehat{\Phi}_{M}\right\Vert $ & $\left\Vert C-\widehat{\Phi}_{C}\right\Vert $ & $\left\Vert F-\widehat{\Phi}_{F}\right\Vert $ & $\left\Vert \widetilde{E}\right\Vert $\tabularnewline
\hline 
Baseline Method \cite{Hart.patil.ea2023} & 3.781 {[}N$\cdot$m{]} & 1.271$\text{[kg\ensuremath{\cdot}m}^{2}\text{]}$ & 6.469 $[\text{kg}\cdot\frac{\text{m}^{2}}{\text{s}}]$ & 6.187 $[\text{N}\cdot\text{m}]$ & -\tabularnewline
Developed Method & 3.030 {[}N$\cdot$m{]} & 1.209$\text{[kg\ensuremath{\cdot}m}^{2}\text{]}$ & 5.060 $[\text{kg}\cdot\frac{\text{m}^{2}}{\text{s}}]$ & 6.032 $[\text{N}\cdot\text{m}]$ & 0.0048 $[\text{kg}\cdot\frac{\text{m}^{2}}{\text{s}}]$\tabularnewline
\end{tabular}
\end{table*}
Simulation results are provided to demonstrate the performance of
the developed SS-LbPINN using a two-link planar revolute robot modeled
in \cite[Eqn. (24)-(26)]{Le.Patil.ea.2022}. Simulations were run
for 50 s with the desired trajectory $q_{d}\triangleq(1-\text{exp}(-0.1))\begin{bmatrix}\frac{3\pi}{8}\mathrm{sin}\left(\frac{\pi}{2}t\right)\\
\frac{3\pi}{8}\text{sin}\left(\frac{\pi}{2}t\right)
\end{bmatrix}\in\mathbb{R}^{2}\text{ [rad]}$, initialized at $q(0)=[0.4,-0.3]^{\top}\text{ [rad]}$ and $\dot{q}(0)=[0,0]^{\top}\text{ [rad/s]}$.
\begin{figure}[h]
\includegraphics[width=1.05\columnwidth]{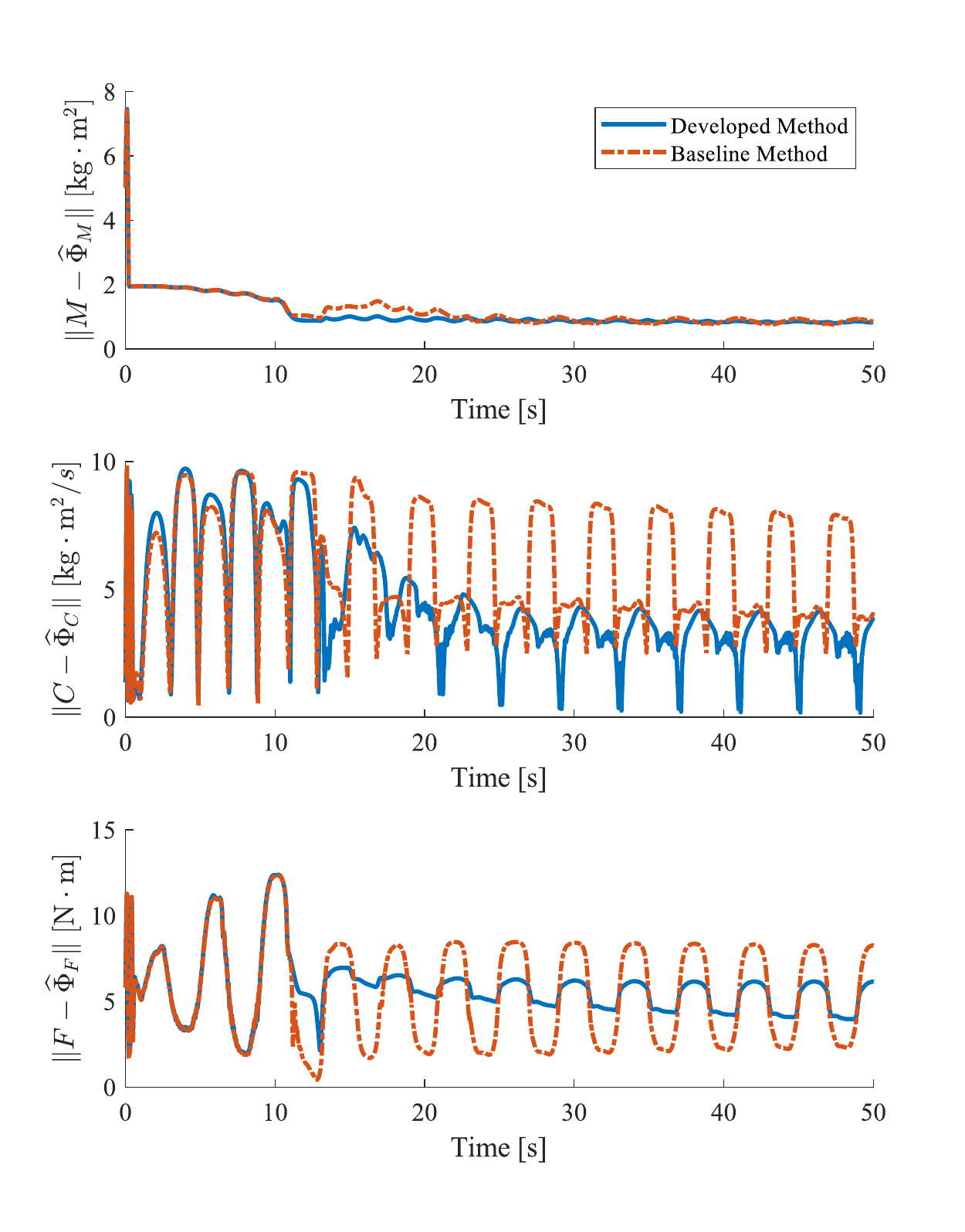}\centering\caption{\label{fig:IndividualFuncApprox}Individual function approximation
errors over time for the developed method and the baseline method
in \cite{Hart.patil.ea2023}.}
\end{figure}
 To examine the effect of noise, white Gaussian noise with signal-to-noise-ratio
(SNR) of 60 dB is added to the position and velocity measurements.
To assess the impact of the skew-symmetric prediction error, a comparison
was done between the developed method and the architecture in \cite{Hart.patil.ea2023}
which does not include the skew-symmetric prediction error. Both architectures
were composed of 3 DNNs with 4 layers, 7 neurons in each hidden layer,
and tanh activation functions for the $M,C,$ and $F$ matrices. The
gravitational effects represented by $G$ were not considered due
to the planar configuration of the robot. The simulations were performed
using the same initial DNN weight estimates which were randomly initialized
from a uniform distribution $U\left(-1,1\right)$. For a fair comparison
between the baseline \cite{Hart.patil.ea2023} and developed architectures,
the gains shared between the architectures were selected to be the
same; because of this, it is expected that both architectures have
similar tracking performance. The control gains are selected as $\alpha=3.8$,
$k_{1}=15.1$, $k_{2}=0.5$, $k_{3}=0.5$, $k_{4}=0.3$ with weight
adaptation gains $\Gamma_{M}=1.1\cdot I_{\varkappa_{M}}$, $\Gamma_{C}=11.5\cdot I_{\varkappa_{V}}$,
and $\Gamma_{F}=9\cdot I_{\varkappa_{F}}$. The gains associated with
the skew-symmetric prediction error were selected as $\gamma_{1}=186.1,$
$\gamma_{2}=1.2$, $\gamma_{3}=6.5$, $\gamma_{4}=2.7$, and $\xi=0.4$.
Both architectures achieve similar tracking performance, with values
of 0.0195 rad under 7.345 Nm of control effort, and 0.0194 rad under
7.350 Nm of control effort for the developed and baseline methods,
respectively. However the SS-LbPINN was motivated by improving the
accuracy of the individual estimates. As shown in Figure \ref{fig:IndividualFuncApprox}\textbf{
}and Table \ref{tab:table} the incorporation of a skew-symmetric
prediction error resulted in a 4.84\%, 21.78\%, and 2.52\%\textbf{
}improvement for the individual matrix approximations for $M,$ $C$,
and $F$, respectively. The improvement in the individual matrix estimations
is reflected in the overall function approximation error (Figure \ref{fig:OveralllFuncApprox})
represented as $\tilde{f}\triangleq\left(M-\widehat{\Phi}_{M}\right)\ddot{q}+\left(C-\widehat{\Phi}_{C}\right)\dot{q}+\left(F-\widehat{\Phi}_{F}\right)$
which exhibited a 19.87\% improvement in the developed architecture
compared to the baseline while requiring similar control effort.

\section{Conclusions}

A physics-informed controller is developed for general uncertain Euler-Lagrange
systems. Leveraging the dynamic structure, the controller is composed
of individual estimates of the unknown terms. Then, leveraging the
individual estimates, an adaptive update law is developed which penalizes
the skew-symmetric terms to exploit the known skew-symmetry of the
system. Unlike traditional NN approaches, the developed SS-LbPINN
control design and update laws are informed by the known physics of
the system. Stability-driven weight adaption laws are developed for
the SS-LbPINN weights in real-time, eliminating the need for offline
pre-training. A Lyapunov-based stability analysis is performed and
guarantees asymptotic convergence of the tracking errors and the skew-symmetric
prediction errors. Simulations validate the SS-LbPINN, showing a 19.87\%
improvement in function approximation and 4.84\%, 21.78\%, and 2.52\%
improvement in $M$, $C$, and $F$ approximations, respectively,
over \cite{Hart.patil.ea2023}, while maintaining similar tracking
error and control effort. \bibliographystyle{ieeetr}
\bibliography{ncr,encr,master}

\end{document}